\documentclass[preprint,12pt]{elsarticle}
\usepackage{amsmath}
\usepackage{amsfonts}
\usepackage{bm}
\usepackage{lipsum}
\usepackage{enumerate}
\usepackage{graphicx}
\usepackage{amsfonts}

\usepackage{mathrsfs}
\usepackage{subfigure}
\usepackage{epstopdf}
\usepackage{url}
\usepackage{verbatim}
\usepackage{amssymb}
\usepackage{hyperref}
\usepackage{color}
\usepackage{listings}

\def\@begintheorem#1#2{\par\bgroup{\scshape #1\ #2. }\it\ignorespaces}
\def\@opargbegintheorem#1#2#3{\par\bgroup%
   {\scshape #1\ #2\ ({\upshape #3}). }\it\ignorespaces}
\def\@endtheorem{\egroup}

\if@onethmnum
  \newtheorem{theorem}{Theorem}
  \newtheorem{lemma}[theorem]{Lemma}
  \newtheorem{corollary}[theorem]{Corollary}
  \newtheorem{proposition}[theorem]{Proposition}
  \newtheorem{definition}[theorem]{Definition}
\newtheorem{example}[theorem]{Example}
\newtheorem{remark}[theorem]{Remark}
\newtheorem{homework}[theorem]{Homework}
\newtheorem{case}[theorem]{}

\else

\fi

\usepackage[T1]{fontenc}

\journal{XXX }

\begin{document}

\begin{frontmatter}



  \title{Critical density triplets for the arrestment of a sphere falling in a sharply stratified fluid\\
    \footnotesize In memory of our friend Bong Jae Chung}
  

\author[1]{Roberto Camassa}
 \ead{camassa@amath.unc.edu}
\author[1]{Lingyun Ding}
\ead{dingly@live.unc.edu}
\author[1]{Richard M. McLaughlin \corref{mycorrespondingauthor}}
\cortext[mycorrespondingauthor]{Corresponding author}
\ead{rmm@email.unc.edu}
\address[1]{Department of Mathematics, University of North Carolina, Chapel Hill, NC, 27599, United States}
\author[1]{Robert Overman}
\author[1]{Richard Parker}

\author[2]{Ashwin Vaidya}
 \ead{vaidyaa@mail.montclair.edu}
\address[2]{Department of Mathematics, Montclair State University, Montclair, NJ, 07043, United States}

\begin{abstract}
  We study the motion of a rigid sphere falling in a two-layer stratified fluid under the action of gravity in the potential flow regime. Experiments at a moderate Reynolds number of approximately 20 to 450 indicate that a sphere with the precise critical density, higher than the bottom layer density, can display behaviors such as bounce or arrestment after crossing the interface. We experimentally demonstrate that such a critical sphere density increases linearly as the bottom fluid density increases with a fixed top fluid density. Additionally, the critical density approaches the bottom layer fluid density as the thickness of density transition layer increases. We propose an estimation of the critical density based on the potential energy. With assuming the zero layer thickness, the estimation constitutes an upper bound of the critical density with less than 0.043 relative difference within the experimental density regime 0.997  $g/cm^{3}$ $\sim $ 1.11 $g/cm^{3}$ under the zero layer thickness assumption.  By matching the experimental layer thickness, we obtain a critical density estimation with less than 0.01 relative difference within the same parameter regime.     

\end{abstract}


\end{frontmatter}

\section{Introduction}
\label{sec:intro}
Stratified fluids are those in which the background, equilibrium density field varies with height. Such systems occur naturally in many environments including lakes, oceans, and the Earth's atmosphere, as well as on other planets.  Sedimentation of particles in stratified fluids ubiquitously occurs in natural environments \cite{magnaudet2020particles} and plays a vital role in marine snow \cite{macintyre1995accumulation,prairie2013delayed},  oil spill properties \cite{adalsteinsson2011subsurface,camassa2016optimal} and distributions of dense microplastics \cite{jamieson2019microplastics} in the oceans, and most recently in marine particulate aggregation \cite{camassa2019first}.

Here, we focus on an interesting phenomenon that occurs when particles cross density interfaces between two fluids of different densities. In work by Abaid et al. \cite{abaid2004internal}, the experimental sedimentation of a sphere in stratified salt water was studied, and an intriguing bounce phenomenon was first documented in which a dense sphere falling in the fluid momentarily stopped, began to rise, before ultimately falling. The momentary levitation of the sphere yields a prolonged settling time, which can contribute to the accumulation of particulate matter in the vicinity of strong density transition layers in the environment, e.g., haloclines or thermoclines \cite{macintyre1995accumulation,widder1999thin,condie1997influence,denman1995biological,deepwell2022cluster}.

We remark that there are three important factors to this bounce phenomenon.  The first parameter is the Reynolds number $\frac{Ua}{\nu}$, where $U$ is a characteristic velocity, $a$ is the radius of the sphere, and $\nu$ is the kinematic viscosity.   Several articles in the literature \cite{camassa2009prolonged,camassa2013retention,camassa2010first} investigated the gravitational settling particles at low Reynolds numbers regime ($Re=0.001$) where a complete first principle based theory is possible.  In the low Reynolds number, no bounce is observed, while the original work by Abaid et al. \cite{abaid2004internal} involved Reynolds numbers in the hundreds.   The second parameter is the relative thickness of the fluid density transition layer $h/a$ which is characterized by the ratio of the layer thickness $h$ to the particle radius $a$.  Several studies  \cite{srdic1999gravitational,verso2019transient} have explored gravitational particle settling in sharply stratified fluids but reported no bounce phenomenon.  In  \cite{srdic1999gravitational} and \cite{verso2019transient}, the parameter $h/a \gg 1$ which takes values $60$ and $20$ respectively, whereas in the work of Abaid et al. this parameter was much smaller, taking values around $3$.  In this paper, we focus on this parameter regime and explore the dependence of the bounce phenomenon on layer thickness experimentally.  
The third factor is the relation between the top fluid density $\rho_{1}$, bottom fluid density $\rho_{2}$, and sphere density $\rho_{b}$.  The sphere rises into the upper fluid when its density is lower than that of the bottom fluid.  When the sphere density is considerably higher than the fluid densities, the sphere penetrates the interface without bouncing back. As a result, predicting the range of sphere densities for which motion reversal is conceivable with known top and bottom fluid densities is intriguing.

Towards that goal, we are interested in using experiments and theory to determine a critical density triplet $(\rho_1,\rho_2,\rho^{*}_b)$ with the constraint $\rho^{*}_b\geq \rho_{2} \geq \rho_{1}$.  For any sphere with the density $\rho^{*}_b \geq \rho_{b} \geq \rho_{2}$, the falling sphere will bounce, if the sphere density equals the critical density, $\rho_b=\rho_b^*$, the falling sphere will just stop momentarily but not rise before ultimately descending to the tank bottom.  Additionally, increasing the sphere density such that $\rho_{b}> \rho^{*}_b$ with fixed $\rho_{1}$ and $\rho_{2}$, or equivalently, decreasing $\rho_2$ with fixed $\rho_{1}$ and $\rho_{b}=\rho^{*}_b$ prevents the sphere from stopping. 
As previously stated, we concentrate on cases with relatively high Reynolds numbers (between 20 and 450, based on the terminal velocities in the bottom and top layers, respectively) and a sharply stratified fluid ($h/a<4$, $h \sim 0.9$ cm, and $a=0.25$ cm).  There are very few studies attempting to estimate these critical densities in the literature.  Perhaps the first attempt was by Camassa et al. \cite{camassa2008brachistochrones}.  In that work they proposed  a coarse criterion for the critical density triplet which is based upon estimating the enhanced buoyancy through an asymptotic calculation of the drift volume induced by a sphere traveling a finite but large distance.
 Here, we aim to improve the critical density estimation based on the potential flow assumption and the system's potential energy for the levitation phenomenon of a sedimenting sphere in such a parameter regime. By analyzing the monotonicity of the potential energy curve, we establish an estimation that depends on the sphere and fluids density, the initial position of the sphere, and the thickness of the fluid density transition layer.  Last, we anticipate this study could have applications in separating particles with different densities.


The paper is organized as follows. In section \ref{sec:setup}, we present the setup of the model and formulate the energy equation of the system. In section \ref{sec:Experiment}, we document the details of the experimental procedure and the critical density obtained by the experimental method. The linear regression of the experimental data shows critical densities satisfy the relation $\rho^{*}_b=1.03 \rho_{2}-0.0295\rho_{1}$. Additionally, we demonstrate that thicker layer transitions are less capable of arresting the sphere.  In section \ref{sec:criticaldensity}, we provide a criterion to estimate the critical sphere density with given top and the bottom fluid density, layer thickness, and the initial position of the sphere. We document the details of our numerical method in appendix \ref{sec:appendix}.

\section{Setup and governing equations}
\label{sec:setup}
\subsection{Setup description}

\begin{figure}
\centering
 \includegraphics[width=0.46\linewidth]{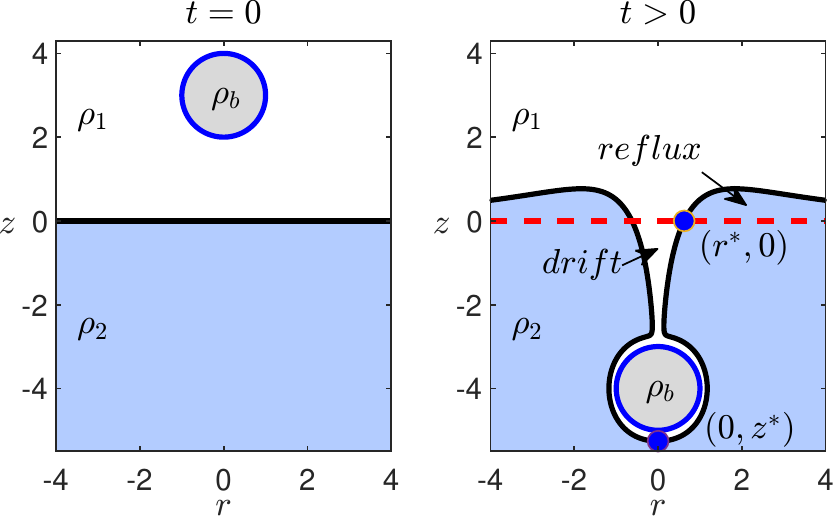}
 \caption{The setup at initial time $t=0$ and a finite time $t>0$. The white, light blue, and gray regions are occupied by the top fluid, bottom fluid, and sphere respectively. The black solid line is the interface between the top and bottom fluid.  }
 \label{fig:sketch}
\end{figure} 

We consider a sphere with the density $\rho_b$ and radius $a$ sedimenting in a two-layer unbounded homogeneous fluid imposed upon the regions above and below an artificial interface  as sketched in figure \ref{fig:sketch}. The top and bottom fluid densities are $\rho_1$ and $\rho_2$ respectively ($\rho_2>\rho_1$).  At time $t=0$, the interface of two fluid layers is centered at $z=0$. The sphere starts at $(0,z_{b} (0))$, $z_{b} (0)= z_0>a$  with an initial velocity $(0,v_{b} (0))$.

Due to the complex nature of the fluid flow around the body, the energy expressions of interest must be determined numerically by evaluating the evolution of the fluid interface. We make the following simplifying assumptions. First, the sphere penetrates the interface very fast, sufficiently so as not to generate any waves. The interface of the two-fluid layers stays sharp at the end of the experiment.  Second, we assume the fluids are inviscid and irrotational. Based on these assumptions, the velocity field induced by the falling sphere can be modeled by the three-dimensional potential flow. To take advantage of the axial symmetry, we adopt the cylindrical coordinate system $r=\sqrt{x^{2}+y^{2}}$. The position $(r(t), z(t))$ of a passive tracer in the fluid satisfies the equation 
\begin{equation}
  \label{eq:potentialFlow3D}
\frac{\mathrm{d} z}{\mathrm{d} t} =   \frac{v_{b} (t) a^3(r^{2}-2\tilde{z}^2 )}{2(\tilde{z}^2 + r^2)^{5/2}} , \quad \frac{\mathrm{d} r}{\mathrm{d}t} = \frac{-3v_{b}(t) a^3\tilde{z}r}{2(\tilde{z}^2 + r^2)^{5/2}},
\end{equation}
where $\tilde{z}= z (t)- \left( z_{0}+\int\limits_{0}^{t}v_{b}(s)\mathrm{d} s \right)$, and $(0,v_{b} (t))$ is the velocity of the sphere. Here, we observe an important property of this model which will greatly simplify the numerical calculation:  the resulting interface shape from passive advection is independent of the time history of the sphere trajectory. To see this, rescaling time via $\int\limits_{0}^{t}v_{b}(s)\mathrm{d} s=k$ in equation \eqref{eq:potentialFlow3D} results in:
\begin{equation}
\begin{aligned}
\frac{\mathrm{d} z}{\mathrm{d} k} =  \frac{-a^3(r^{2}-2\tilde{z}^2 )}{2(\tilde{z}^2 + r^2)^{5/2}} , \quad \frac{\mathrm{d} r}{\mathrm{d}k} = \frac{-3 a^3\tilde{z}r}{2(\tilde{z}^2 + r^2)^{5/2}},\quad \tilde{z}=z-(z_{0}+k),
\end{aligned}
\end{equation}
which is the equation of the tracer in the case that the sphere moves with the unit speed with respect to the pseudo time $k$. Thus, the interface resulting from any sphere motion ending at the same position is identical.  There are two options to explore here.  First, one could explore the consequences of employing energy conservation to self-consistently evolve the sphere and fluid under the assumptions of potential flow.  Second, one could study the potential energy stored in the fluid as a function of the three densities through a sphere moving at a constant speed.  We will study the latter option here in this paper because of the independence of path history and its direct theoretical implications, and discuss the limitations of the former in the conclusion section.  Without loss of generality, we assume the sphere has a constant speed in the numerical simulation.

Last, we assume the fluid density linearly depends on the concentration of the solute, for example, the sodium chloride solution \cite{hall1924densities}. Since the solute is passively advected by the fluid flow and the diffusion is negligible in the experimental time scale,  the fluid density field satisfies the advection equation
\begin{equation}\label{eq:advectionEquation}
\partial_{t} \rho + \mathbf{u}\cdot \nabla \rho=0,
\end{equation}
where $\mathbf{u}$ is the velocity field provided in equation \eqref{eq:potentialFlow3D}. In this study, we consider the following initial density profile, 
\begin{equation}\label{eq:InitialDensityProfile}
\rho (r,z,0)=\rho_{I} (z)=
\begin{cases}
  \rho_{1} & \frac{L_{\rho}}{2}\leq z \\
\rho_{1}+ \frac{2 (\rho_2-\rho_{1})}{L_{\rho}^{2}} \left(\frac{L_{\rho}}{2}-x\right)^2 \left(\frac{x}{L_{\rho}}+1\right)& -\frac{L_{\rho}}{2}<z < \frac{L_{\rho}}{2}\\
 \rho_{2} & z\leq -\frac{L_{\rho}}{2}
\end{cases},
\end{equation}
which has a continuous first order derivative. As $L_p\rightarrow 0$, equation \eqref{eq:InitialDensityProfile} converges to the step function 
\begin{equation}
\rho (r,z,0)=
\begin{cases}
  \rho_{1}& 0\leq z\\
  \rho_{2} &  z<0\\
\end{cases} .
\end{equation}

\subsection{Non-dimensionalization}

We non-dimensionalize the equations and formulae via the following change of variables
\begin{equation}
\begin{aligned}
  & ar'=r,\quad  az'=z,\quad \frac{a}{U} t'=t,\quad a^{4} g \rho_{1}P'=P, \quad \rho_{1}\rho'= \rho, \quad aL_{\rho}'=L_{\rho} ,\quad
  \mathrm{Re}= \frac{Ua}{\nu}.
\end{aligned}
\end{equation}
We can distinguish the dimensional and dimensionless variables by the units after their values. The variables in section \ref{sec:Experiment} are in the dimensional form. Without specification, the variables in the next subsection and  section \ref{sec:criticaldensity} are dimensionless. Hence, we can drop the prime without confusion. The non-dimensionalized initial density profile is
\begin{equation}\label{eq:InitialDensityProfile Nondimensional}
\rho (r,z,0)=\rho_{I} (z)=
\begin{cases}
1& \frac{L_{\rho}}{2}\leq z \\
1+ \frac{2 (\rho_2-1)}{L_{\rho}^{2}} \left(\frac{L_{\rho}}{2}-x\right)^2 \left(\frac{x}{L_{\rho}}+1\right)& -\frac{L_{\rho}}{2}<z < \frac{L_{\rho}}{2}\\
 \rho_{2} & z\leq -\frac{L_{\rho}}{2}
\end{cases}.
\end{equation}

\subsection{The potential energy}

Our goal is to use the energy of the sphere-fluid system to capture the arrestment of the spherical body as it moves through the two fluids. A full scale dynamic explanation of the phenomena is very complex. We believe that our explanation provides an alternative and simpler explanation for the levitation phenomenon.

 The total mechanical energy of the system at any instant of time can be given by
 \begin{equation}
   \label{eq:totalenergy}
E(t) = P_1(t) + P_2(t) + P_b(t) + K_b(t) + K_f(t),
\end{equation}
where $P_1(t)$ and $P_2 (t)$ are the potential energies of the top and bottom fluids respectively. $P_b (t)$ is the potential energy of the body. $K_b(t)$ and $K_f(t)$ are the kinetic energies of the body and fluid respectively. It is possible to express the kinetic energy of the fluid in terms of the added mass \cite{eames1994drift,taylor1928energy}.

First, we consider the extremely sharp stratification, namely, $L_{\rho}=0$. 
A consequence of the law of conservation of mass is that the sphere penetrating into the bottom fluid causes a displacement of the top layer into the bottom and vice versa, the bottom fluid into the top layer.  Therefore in estimating the change in potential energies, the exact shape of the interface at a given time instant  must be known. It is possible to write $\Delta P_{1}=P_{1} (t)-P_{1} (0)$ as a result of gained volume by the top fluid in the drift region and a lost volume in the reflux region. Similarly, $\Delta P_{2}=P_{2} (t)-P_{2} (0) $ can be thought of as potential energy of the bottom fluid due to a volume lost in the drift region and gained volume in the reflux region. We can therefore think of the net change in potential energy of the entire fluid as coming only from the drift and reflux volume regions in addition to the gravitational potential energy contribution of the fluid displaced by the body. The change of potential energy can then be written in the form,
 \begin{equation}\label{eq:DeltaPe}
   \begin{aligned}
     \Delta P&=  \Delta P_1 + \Delta P_2+\Delta P_{b}  \\
     &=   \pi(\rho_2-1) \left( \int_{r^{*} (z_{b})}^{\infty} r z(r)^2  \mathrm{d} r- \int_{z^{*} (z_{b})}^{0} r(z)^2 z  \mathrm{d} z \right) + \frac{4\pi}{3}(\rho_{b}  -1) (z_{b} (t) -z_{b} (0)),
\end{aligned}
\end{equation}
where $(r^{*} (z_{b}),0)$ are the coordinates of the point of zero Lagrangian displacement, $(0,z^{*} (z_{b}))$ is the lowest point on the interface (see figure \ref{fig:sketch}). The asymptotic expansion of $\Delta P$ in the limit of some parameters is available via the asymptotic expansions provided in \cite{camassa2008brachistochrones,lighthill1956drift,yih1985new,yih1997evolution}.  In other cases, we have to compute $\Delta P$ numerically.

\begin{figure}
  \centering
    \includegraphics[width=1\linewidth]{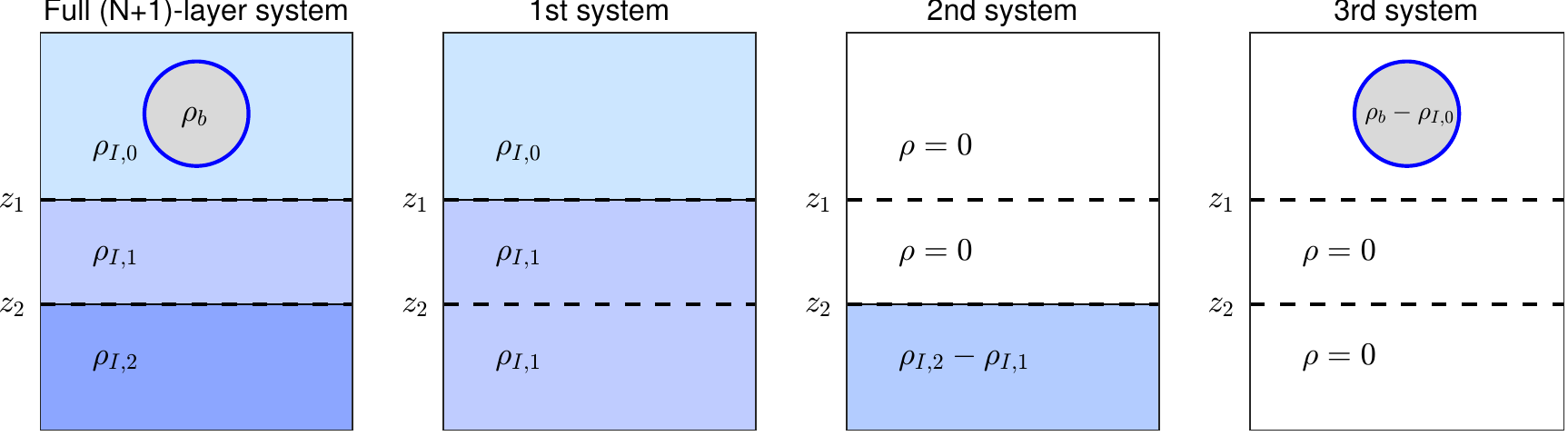}
  \hfill
  \caption[]
  {A schematic of decomposing the full $(N+1)$-layer system into $N$ two-layer systems and a system that only contains the sphere. The potential energy in the full system equals the sum of the potential energy in all subsystems. Here $N=2$, $t=0$, $\rho_{I,0}=1$, $\rho_{I,N}=\rho_{2}$. $z_{n}$ is the height of the $n$-th interface. }
  \label{fig:plotProofSketch}
\end{figure}

  Second, for the case with a non-zero density transition layer thickness, we prefer to use the results in the zero layer thickness case rather than solve the full advection equation \eqref{eq:advectionEquation}. We consider $N$ artificial interfaces at $z=z_{i}$ which satisfy $-\frac{L_{\rho}}{2}=z_{N}<z_{N-1}<\hdots<z_{2}<z_{1}= \frac{L_{\rho}}{2}$. The fluid density between $n$th and $(n+1)$th layer is approximated by the density at the middle point $\rho_{I,n}=\rho_{I} ((z_{n+1}+z_{n})/2)$. Since the density field is passively advected by the flow, we can divide the $(N+1)$-layer system into $N$ independent two-layer systems while conserving the total potential energy (see the schematic in \ref{fig:plotProofSketch}).  The first system consists of two fluids separated by a sharp interface located at $z=z_1$. The top fluid density is $1$ and the bottom fluid density is $\rho_{I,1}$. The interface in $n$th system ($n>1$) is located at $z=z_n$. The top and bottom fluid density are  $0$ and $\rho_{I,n}-\rho_{I,n-1}$, respectively. The $(N+1)$th system only contains a sphere centered at $z_{b}$ with the density $\rho_{b}-1$. Clearly, the summation of the potential energy of these $N+1$ systems equals the potential energy of the original system. Since each system has a  two-layer stratified fluid, we can apply the previous conclusion \eqref{eq:DeltaPe} and obtain the following expression of the change in potential energy
\begin{equation}\label{eq:DeltaPeFiniteThickness}
\begin{aligned}
  &\Delta P=\frac{4\pi}{3}(\rho_{b}  -1) (z_{b} (t) -z_{b} (0))\\
  &+\lim\limits_{N\rightarrow \infty}  \sum\limits_{n=1}^N  \pi \left(
  \rho_{I,n}-\rho_{I,n-1} \right) \left( \int_{r_{n}^{*} (z_{b})}^{\infty} r z_{n}(r)^2  \mathrm{d} r- \int_{z_{n}^{*} (z_{b})}^{0} r_{n}(z)^2 z  \mathrm{d} z \right),
   \end{aligned}
\end{equation}
where $r_n (z)$, $z_{n} (r)$, $r_{n}^{*} (z_{b})$, $z_{n}^{*} (z_{b})$ are associated with the interface starts at $z=z_{n}$ and $\rho_{I,0}=1$, $\rho_{I,N}=\rho_{2}$. In the numerical simulation, we distribute $z_{n}$ uniformly and $N = 30 \sim 40$ is enough to obtain desirable results.

\section{Experimental methods and results}
\label{sec:Experiment}
Our experimental study involved dropping several spherical beads into a tank containing density stratified liquids by varying the fluid and sphere densities and the layer thickness. In following sections, we detail the exact procedure followed for various aspects and stages of the experimental study. 

\subsection{Tank, bath and camera setup}


\begin{figure}
  \centering
    \includegraphics[width=0.6\linewidth]{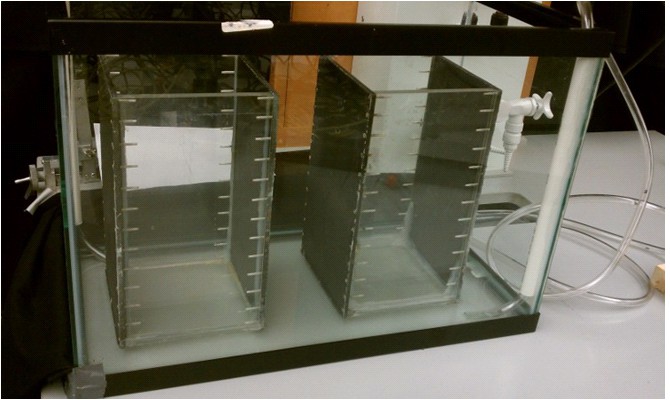}
 
  \hfill
\caption{Setup with two inner experimental tanks and outer tank as a thermal bath which is connected to a recirculating thermally controlled reservoir.}
  \label{fig:tankBath}
\end{figure}

As presented in figure \ref{fig:tankBath}, the setup consists of two experimental tanks placed within a thermal bath. The outer tank (thermal bath tank) is regulated by a Thermo Scientific Neslab RTE-7 Digital Plus Refrigerated bath. The thermal bath is maintained at 19 degrees Celsius while taking data. Each of the inner experimental tanks consists of two Plexiglas sides for ease of viewing, a Plexiglas bottom, and two sides made of copper plates to assist in thermalization. The copper plates are coated in a protective sealant to prevent corrosion. The bottom layer of fluid is prepared in one of the inner tanks and the top layer in the other. The tank with the bottom layer of fluid is filled only halfway as the experiment will be run in this tank after pouring the top layer. The outer tank is a glass fish tank for ease of viewing. The Neslab machine is connected to the fish tank via flexible PVC tubing. Two solid one-foot sections of PVC pipe were glued into opposing corners of the fish tank. The flexible tubing is run over the top of the tank and down through these PVC pipes so that the input and output flow can be placed parallel to the experimental tank sides in an effort to minimize any vibrations upon the experimental tanks.

As stated earlier, there are two inner experimental tanks where salt water solutions are prepared, and one outer tank for the thermal bath. The two inner tanks have the outer dimensions of 7" by 7" by 12.5". The two copper plates are each 0.25" thick, and the two Plexiglas sides and bottom are each 0.5" thickness. This results in inner dimensions of 6.5" by 6" by 12". The thermal bath tank was simply a standard glass aquarium measuring 24.5" by 12.5" by 16.5". On the rear of the outer tank behind the experimental tank in which the bead is dropped in, a background is placed. A variety of backgrounds were used. For the human viewer, a black and white checkerboard pattern of $0.635$ cm by $0.635$ cm squares produced the best visualization of the layer transition. The compression of the squares is easy to discern, and this gives a very clear view of the layer. For the purposes of tracking the bead in the DataTank script, a checkerboard pattern with 1 mm by 1 mm squares produced cleaner data. A solid background was never used, but this would most likely produce data with even less noise in the script, but such a background makes it much more difficult to discern a clean transition with the human eye. 

\begin{figure}
  \centering
  \subfigure[]{
    \includegraphics[width=0.46\linewidth]{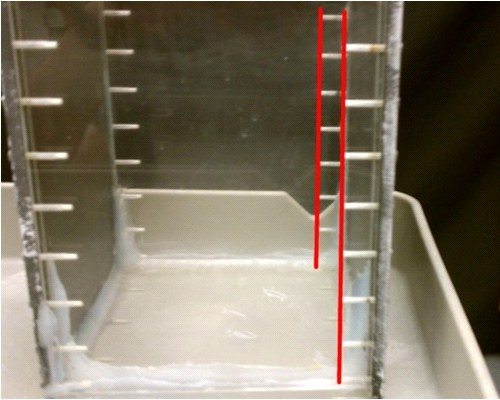}
  }
    \subfigure[]{
    \includegraphics[width=0.46\linewidth]{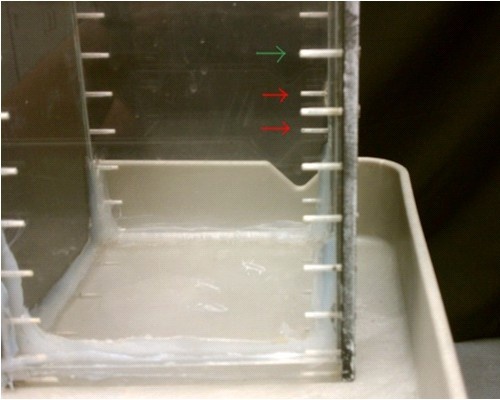}
  }
  
  \hfill
  \caption{\textbf{(a)} Vertical alignment: the tip of the screw holes on the back wall should be inside the top of the screw holes on the front wall. Also the distance between the line of tips on both the back wall and the line of tips on the front wall (lines shown in red) should be the same for both the left and right side of the tank.  \textbf{(b)}  The green arrow shows good alignment horizontally. The back screw hole is directly behind the front screw hole. Red lines show these screw holes are not aligned with the screw hole on the front of the tank. 
  } 
  \label{fig:tankAlignment}
\end{figure}
\begin{figure}
  \centering
    \includegraphics[width=0.23\linewidth]{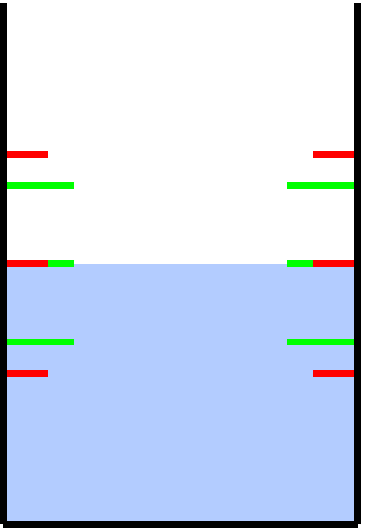}
  \hfill
\caption{Optimal setup: Green lines are screw holes on back wall, and red are those on the front. The distance between the hole ends on the front and back of the tank (red and green liens) are equal on both the left and right side of the tank. Also, at the level of the water-air interface (before top layer is poured) the rear screw holes are directly behind those of the front. Both above and below the water-interface line the screw holes on the back wall appear before those of the front wall. They are ``inside'' the next set of screw holes on the front wall, and the difference in height between this set of red and green lines is equal for all four such sets. } 
  \label{fig:TankScrewHole}
\end{figure}

A Sony HD camcorder is used for the duration of filming on the project. The camera is setup a meter in front of the inner tank the experiment will be run in. The camera is leveled to be on alignment with the water-air interface in the tank which contains the bottom layer. The alignment is performed using the lines on the front and back of tank. At the interface level, the lines on the front pane of the tank should be directly in front of the lines on the rear pane of the tank. Also the lines on the rear of the tank should stick out from the end of the lines on the front of the tank by equal amounts on both the left and right sides of the tank at the interface level (see figure \ref{fig:tankAlignment} and \ref{fig:TankScrewHole}). After completing the alignment phase, a meter stick is put in the center of the tank where the bead will fall and the camera is focused on the smallest demarcations on the stick.

\subsection{Stratification setup}

The density setup begins by filling the tanks with deionized water. Diamond Crystal Extra Coarse Solar Salt is then added to the water to bring the solution up to the desired density. An aquarium fishnet is used to hold the salt in the tank while mixing. The salt dissolves faster using the fishnet since there is a greater surface area of salt exposed to the water as opposed to being piled on the bottom. Also having the salt in the fishnet allows quick removal of the salt to avoid overshooting the target density. While fixing the density, the temperature must be maintained at 19 degrees Celsius. Both solutions (top and bottom layer) are brought up to the desired densities in separate tanks. To ensure the solution is fully mixed, once attaining the target density, another sample is tested to ensure the density is true.

Pouring the layer is the most delicate part of the experiment as the two salt water solutions will mix very easily. Care must be taken to pour the layer carefully and to not bump the tanks while the layer is being poured. The layer is poured through a diffuser. The diffuser is simply a combination of two types of foam. The porous, sponge-like center allows the layer being poured to slowly settle on top of the bottom layer. The outer foam (blue Styrofoam insulation board) keeps the diffuser buoyant enough to float up as the water level rises. Before using the diffuser, the diffuser must be primed with the top layer solution. For this reason the tank with the top layer solution should be filled to the top even though the tank with the bottom layer is only filled half way. If the diffuser is primed with deionized water and the top layer has a density different from deionized water, then as the top layer is poured through the diffuser, there will be a strong tendency for the density to drift away from the original density as it flows through. Therefore the diffuser needs to be primed with a few cups of the same solution which will eventually be poured through the diffuser. Along these lines the diffuser must also be cleaned with deionized water after use to prevent salt accumulation. After attaining both desired densities and priming the diffuser, the diffuser is placed on the top of the experimental tank, floating on top of the bottom layer. A syringe is then used to gently pour the top layer through the diffuser. The general idea is to pour very slowly at the start (a rapid drip from a syringe) and then speed up as the distance between the diffuser and interface layer increases. For a more quantifiable rate, using 60 ml syringes with the exit hole widened to 8 mm the first pour through the diffuser should take approximately 45 s. The final pour through the diffuser should take 1.5 s. At the start, going too quickly will mix the interface, resulting in a poor transition layer. Once the diffuser sits an inch or so above of the interface, going too slowly just allows the interface more time to diffuse. The goal when pouring the layer for this experiment is to keep the layer thickness (see Layer Profiling) under 1.00 cm. When the tank is filled to just shy of the top, carefully lift the diffuser straight up, keeping it level while doing so to prevent water from pouring suddenly from it. Any big drips or sudden movements while removing the diffuser will disturb the layer. The purpose of filling the experimental tank all the way up is to ensure that the bead has enough time to reach its terminal velocity in the solution before encountering the transition layer.

	
The beads dropped are made of glass and have diameters of 4-5 mm. The beads have a very slight peak on one end as a result of manufacturing. Although very slight, this little extra glass makes this point the heaviest part of the bead. Therein, when dropping the bead, care is taken to orientate the bead so this point is on the bottom. Otherwise the bead will spin in an effort to orientate itself in this manner as it falls. Before releasing the bead but while holding the bead under the surface of the water, care is taken to remove any air bubbles adhering to the bead by rolling the bead between two fingers.  The beads were manufactured by American Density Materials.  To accurately measure their precise density, we used bisection search with Archimedes method using different tanks of salt water in insulated containers.  Fluids densities were accurately measured using an Anton-Parr DMA 4500 Densitometer.  

Layer profiling is performed using a conductivity probe attached to a Velmex, Inc. high precision UniSlide. The layer is profiled after the bead has been dropped because the probe disturbs the layer upon passing through it. Once the slider is clamped to the tank to ensure it doesn't move while the probe is being lowered, measurements of conductivity and temperature are taken at 0.1cm increments beginning at approximately a centimeter above the interface and continuing to approximately a centimeter below (until the readings level off). The layer thickness is calculated as follows. After all readings have been taken, the conductivity and temperature readings are converted to densities using previous experimental tables for salt water solutions. Layer thickness is quantified by focusing on the change from the lowest density solution (top fluid) to the highest density fluid (bottom fluid). The layer thickness is measured as the distance between the points of 10\% and 90\% change in density.  More precisely, We define $L_{10}$ to be the height at which the density profile takes the value of the top density plus $10$ percent of the total density variation, and similarly define $L_{90}$ to be the height at which the density is top density plus $90$ percent of the total density variation. Then the Layer thickness is given by $L=|L_{10}-L_{90}|$.  

For the entirety of the experimental data (except for the section on the effects of diffusion time on layer thickness) the layer thickness was kept under 1.1cm. The majority of the runs had a layer thickness of 0.85 cm- 0.9 cm. 
The final step in the experiment is to film a meter stick in the tank, right where the bead fell. This is used for attaining a scale in the script which calculates the minimum velocities.

\begin{figure}
\label{ef5}
\centering
 \includegraphics[width=60mm]{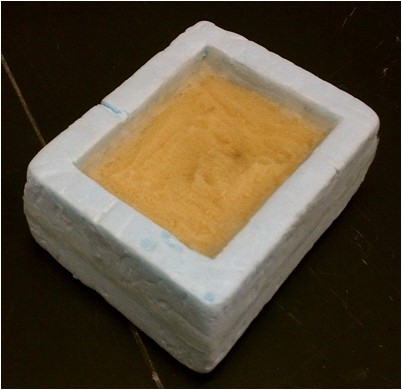}
\caption{The diffuser for pouring the stratified fluid.} 
\end{figure}

\subsection{Experimental results}

We repeated the experiment with hundreds of combinations of the various sphere and fluid densities to experimentally search for the critical density triplet $(\rho_1,\rho_2,\rho^{*}_b)$ as defined in section \ref{sec:intro}.  This is an extremely labor-intensive task as each measurement for one bottom density requires preparing the salt solution with the desired density, pouring an entirely fresh layer, and measuring the density profiles, which takes hours.  With given top fluid and sphere densities, to find one critical bottom density takes at least 10 independent fresh tanks.

\begin{table}[t]
\centering
\begin{tabular}{|c|c|c|c|c|}
\hline
Bead 	&Top 	&Bottom &	Min  &	Layer \\
density(g/cc) & density(g/cc) & density(g/cc) & velocity (cm/s) & thickness(cm)\\
\hline				
1.0901	&0.997	&1.08680&	0.069&	0.814641521\\
$\pm$	&0.997&	1.08678&	0.055&	0.830600263\\
0.0001	&0.997&	1.08683&	0.065&	0.874706739\\
\hline				
1.07495&	0.997&	1.07166&	0.056&	0.843531654\\
$\pm$	&0.997&	1.07182&	0.001&	0.903259318\\
0.0001	&0.997&	1.07170&	0.039&	0.904773401\\
\hline
1.05018&	0.997&	1.04805&	0.029&	0.819058413\\
$\pm$	&0.997&	1.04805	&0.052&	0.817808592\\
0.0002	&0.997&	1.04803&	0.098&	0.840865709\\
\hline				
1.03997&	0.997&	1.03761&	0.018&	0.896928129\\
$\pm$	&0.997&	1.03759	&0.06	&0.795575335\\
0.0002	&0.997&	1.03761&	0.051&	0.905041625\\
\hline				
	1.03506&	0.997&	1.03335&	0.006&	0.885285341\\
$\pm$	&0.997&	1.03335&	0.013	&0.87707262\\
0.00015	&0.997&	1.03333&	0.052&	0.850024643\\
\hline				
	1.02017	&0.997&	1.01886&	0.048	&0.874341231\\
$\pm$	&0.997&	1.01882&	0.035&	0.882575525\\
  0.0001	&0.997	&1.01883&	0.067&	0.890235316\\  
  \hline
1.0901&	1.02998&	1.08725&	0.02&0.959324843	\\
$\pm$ &	1.03006&	1.08726&	0.01&0.950414969	\\
0.0001&	1.03000	&1.08728&	0.05	&0.98831011\\
  \hline  
1.0901&	1.03999&	1.08755	&0.04&	0.955474843\\
$\pm$   &	1.04001&	1.08755&	0.05&	0.940466969\\
0.0001&	1.04000&	1.08755&	0.06&	0.99331213\\
  \hline				
\end{tabular}
\caption{ Critical density triplet $(\rho_1,\rho_2,\rho^{*}_b)$ and related experimental parameters. The fourth column is the minimum speed during the whole falling process. The radius of the spheres is 0.25 cm.  }
  \label{tab:ExperimentalResults}
\end{table}

 Table \ref{tab:ExperimentalResults} shows the critical density triplet $(\rho_1,\rho_2,\rho^{*}_b)$ and related experimental parameters. We calculate the sphere speed by a DataTank script. Because of the spatial and temporal resolution, as well as camera noise, the speed will never be exactly zero. Hence, we report the minimum speed in each experiment and adopt a consistent criterion to determine the arrestment. We have two observations from table \ref{tab:ExperimentalResults}. First, from the first six rows in the table, we see that $\rho_{2}$ increases as $\rho_{b}$ increases when $\rho_1$ is fixed. Second, from the first row and the last two rows in the table, we see $\rho_2$ slightly decreases as $\rho_1$ increases with a fixed sphere density. More interestingly, the linear regression yields the following formula between $\frac{\rho^{*}_b}{\rho_{1}}$ and $\frac{\rho_{2}}{\rho_{1}}$, 
\begin{equation}\label{eq:LRExperiment}
  \begin{aligned}
\frac{\rho^{*}_b}{\rho_{1}}=  a_{1}\frac{\rho_{2}}{\rho_{1}}+a_{2},
\end{aligned}
\end{equation}
where $a_1=  1.03$, $a_2= -0.0295$ with 95\% confidence bounds, respectively,  $(1.019, 1.042)$ and    $(-0.04183, -0.01718)$.  The summed square of residuals is 4.5369e-07 and the R-square value is  0.99987, which indicates a strong statistical linear relation. One can also see the linear relation in  figure \ref{fig:CompareExperiment} which will be elaborated in the next section.


A brief inquiry as to the effects of diffusion time on the resulting layer thickness and critical density was carried out. The purpose of this was mainly to convince those carrying out the experiment that a difference in layer thickness between say 0.85 cm and 0.95 cm would not drastically distort the data. This is important because of the difficulty of ascertaining a precise and repeatable layer thickness every time. The process was simply pouring a tank and waiting for a certain amount of time before dropping the bead and measuring the layer thickness. All runs were done with the same bead of the density of 1.03997 g/cc. The critical density found with de-ionized water on top and no wait time between finishing pouring of the layer and dropping the bead was 1.03761g/cc with a layer thickness of 1.0072 cm. With a two-hour wait time, the critical density was found to be 1.03770 g/cc with a layer thickness of 1.4998 cm. With a three-hour wait time, the critical density was 1.03772 g/cc with a layer thickness of 1.6747 cm. This shows that after three hours of diffusion, the critical density is shifted by about 0.0001 g/cc. Hence this provides some comfort for the few minutes of difference in time pouring the layer for each run, although more work on this could be done.

 We could obtain a rough estimation of layer growth rate by assuming the density profile is $\rho (z)=\rho_{1}+ \frac{\rho_{2}-\rho_{1}}{2} \left(\text{erf}\left(\frac{-z}{2 \sqrt{\kappa  (t+t_{0})}}\right)+1\right)$, where $\text{erf} (z)= \frac{2}{\sqrt{\pi}} \int\limits_0^z e^{-t^2}\mathrm{d} t$. The 10-th and 90-th  percentile of $\text{erf} (z)$ are around $-0.90619$ and $0.90619$, respectively. Based on our definition of the density transition layer thickness, we have $L=3.62478 \sqrt{\kappa  (t+t_{0})}$. Applying the linear regression on $L^{2}$ with respect to $t$, namely, $L^{2}=a_{1}t+a_{2}$, we have $a_{1}=0.0001666$, $a_{2}= 1.023$ with 95\% confidence bounds $ (0.0001122, 0.0002209)$ and $(0.6159, 1.431)$, respectively.  The summed square of residuals is 0.0011 and the R-square value is  0.9993. Comparing these two expressions of the layer thickness $L$, we obtain $t_0=6445.87$ s and $\kappa=1.197809\times 10^{-5}$ $cm^{2}/s$. The molecular diffusivity $\kappa$ computed here is close to the diffusivity of NaCl reported in the literature \cite{vitagliano1956diffusion} which is around $1.3\times 10^{-5} \sim 1.6 \times10^{-5}$ $cm^{2}/s$. 

It is worth noting that because of the intrusive manner of layer profiling with a probe, each of these data points was from separate runs. Since each of these runs was from separated pours, the starting layer thickness differed between the three.

\section{Critical density and  energy criterion}
\label{sec:criticaldensity}
Camassa et al.  \cite{camassa2008brachistochrones} proposed that for a reversal of motion to occur, the averaged density of the sphere and drift fluid must necessarily be less than the density of the bottom layer fluid, in order to have negative buoyancy in the system, which leads to a coarse criterion for the critical density triplet $\bar{\rho}_{b}= (1+c)\rho_{2}-c\rho_{1}$, where $c$ is the ratio of drift volume to the sphere volume. In the case that sphere travels from positive infinity to negative infinity, $c= \frac{1}{2}$. This criterion also shows a linear dependence between the critical densities which agrees with the experimental observations. If $\rho_1=0.997$ g/cc, $\rho_2=1.0376$ g/cc, table \ref{tab:ExperimentalResults} shows $\rho_{b}^{*}=1.04$ g/cc, while this criterion predicts $\bar{\rho}_b=1.0579$ g/cc, which has the relative difference $\frac{\bar{\rho}_{b}-\rho_{1}}{\rho_{b}^{*}-\rho_{1}}-1\approx 0.4163$. Considering that the sphere travels a finite distance, $c$ could be smaller at roughly $0.44$. Therefore this reduces the relative difference to $0.35$, which is still a relatively large difference. Of course, such large errors are to be expected as applying the drift volume directly to the sphere's buoyancy is at best a coarse approximation.

In this section, instead of considering the drift volume, we aim to improve the critical density estimation based on the system's potential energy. With numerical simulations, we next show that the potential energy as a function of the sphere position can change from a non-monotonic to a monotonic function of position as the sphere density increases. This observation will provide a criterion to estimate the critical sphere density such that any sphere with a density higher than the critical value cannot arrest. Since we have non-dimensionalized the problem, we set the radius and sphere speed to be unity in the numerical simulations below.

\begin{figure}
  \centering
  \subfigure[]{
    \includegraphics[width=0.46\linewidth]{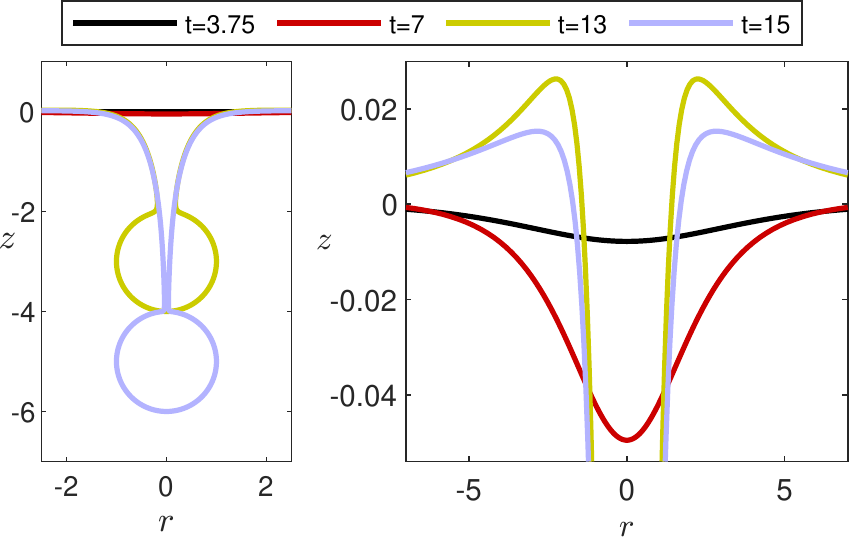}
    }
    \subfigure[]{
    \includegraphics[width=0.46\linewidth]{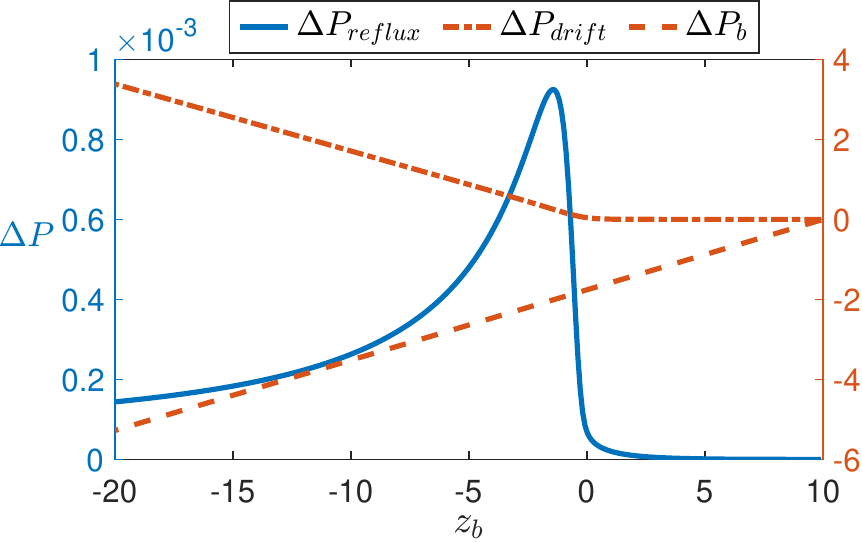}
  } 
  \hfill
  \caption[]
  { \textbf{(a)} The figure shows the time evolution of the interface based on the numerical simulation,  as the sphere falls from the upper layer to the lower one. The drift region is very apparent in the left panel.  The right panel shows a zoomed in view of the reflux region at different times.  \textbf{(b)} The dual $y$-axis chart shows, in the same simulation as panel (a), the change of potential energy contributed by the reflux region $\Delta P_{\mathrm{reflux}}$ {\color{blue}(left axis)}, drift region $\Delta P_{\mathrm{drift}}$ {\color{red}(right axis)} and body $\Delta P_{b}$ {\color{red}(right axis)} when the center of the body $z_{b}$ at different positions. Notice the scale difference in $y$-direction. The parameters are  $\rho_{2}=1.04$, $\rho_{b}=1.042$, $z_{0}=10$. }
  
  \label{fig:InterfaceEnvolutionRefluxDriftPotential}
\end{figure}
Figure \ref{fig:InterfaceEnvolutionRefluxDriftPotential} (a) shows a typical interface evolution as the sphere falls. The scale of the reflux region is small compared to the drift region. Figure \ref{fig:InterfaceEnvolutionRefluxDriftPotential} (b) shows the variation of the change in potential energy contributions due to the sphere, the reflux and drift regions as indicated in equation \eqref{eq:DeltaPe}. The contribution due to the drift region can be seen to far exceed that due to the reflux region as would be expected in free space. After all, the drift volume is carried to infinity along with the moving sphere. While the drift contribution increases monotonically in the range of values computed, the reflux contribution peaks as the sphere just cross the initial interface and then decays.

\begin{figure}
  \centering
    \includegraphics[width=1\linewidth]{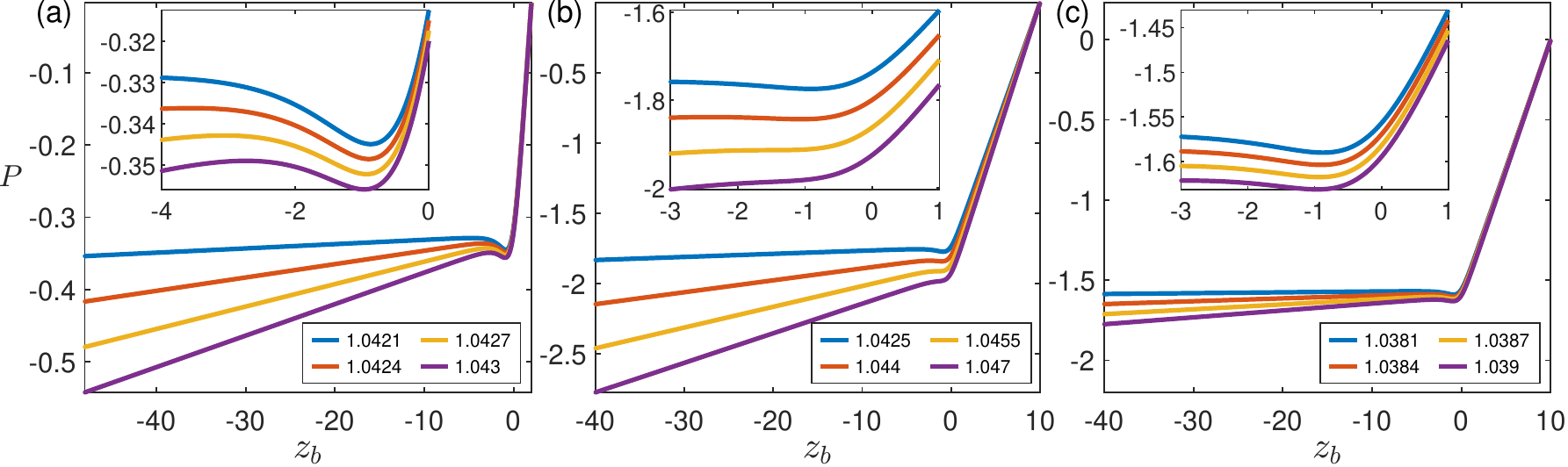}
  \hfill
  \caption[]
  {The figure denotes the net potential energy variation in the entire system for different parameters. The parameters are  $\rho_2=1.042$  in panel (a, b), $\rho_2=1.038$ in panel (c), $z_{b} (0)= z_{0} =2$ in panel (a), and $z_{0}=10$ in panel (b, c). The legend shows the corresponding value of $\rho_{b}$ for each curves. The insets are zoomed in versions of each pictures. }
  \label{fig:PotentialDifferentDensity}
\end{figure}

In figure \ref{fig:PotentialDifferentDensity} we plot the total potential energy versus $z_{b}$ with  $\rho_2=1.042$  in panel (a, b) and $\rho_2=1.038$ in panel (c) while varying the sphere density $\rho_b$. The insets are zoomed-in view of the original picture near the critical points. The different columns correspond to different starting points for the sphere, namely $z_{b} (0)= z_{0} =2$ in panel (a), $z_{0}=10$ in panel (b, c). Figure \ref{fig:PotentialDifferentDensity} is very telling; as $\rho_b$ is varied we see significant variations in the types of curves produced: (i) The early part of each of these curves begins with the potential energy of the system decreasing with $z_b$ caused by the falling sphere whose potential energy decreases. The potential energy contributions of the fluid are non-existent at this stage, (ii) the next phase of this curve occurs when the sphere reaches the interface. The sphere's potential energy continues to decrease however the drift and reflux regions now take place, resulting in a positive contribution to the potential energy which could counter the negative values of the sphere.  The density of the sphere now becomes important henceforth. For the case that $\rho_b$ is slightly larger than $\rho_2$ as is seen clearly in figures \ref{fig:PotentialDifferentDensity} (a, c), the positive energy of the fluid wins resulting in a minimum in the energy curve. Then the energy curve rises briefly as the sphere penetrates the interface and soon after begins to fall again as the fluid's potential energy fails to overcome that of the sphere. For the case when $\rho_b$ is sufficiently larger than $\rho_2$, we see that the energy curve can be completely dominated by the potential energy of the sphere which shows monotonically decreasing behavior.
Therefore, we denote the sphere density where the transient between these two cases happens as $\bar{\rho}_b$, which provides an estimation for $\rho^{*}_b$ in the critical density triplet. 

Mathematically, when the sphere density reaches the critical value, $\bar{\rho}_b $, there exists a degenerate critical point on the curve, $z_b=z_b^*$, such that 
\begin{equation}
  \label{eq:DegenerateCritialPoint}
\left. \partial_{z_{b}} P \right|_{z_b=z_b^* }  = \left. \partial^{2}_{z_{b}} P \right|_{z_b=z_b^* }  = 0.
\end{equation}
Equivalently, $\partial_{z_{b}} P$ is non-positive for $z_{b}<z_{0}$ and only equals zero at one point $z_b=z^{*}_{b}$. Next, we first consider the density profiles with the zero density transition layer thickness and then study the profile with non-zero layer thickness.  

\subsection{Zero density transition layer thickness}
\begin{figure}
  \centering
      \subfigure[]{
    \includegraphics[width=0.46\linewidth]{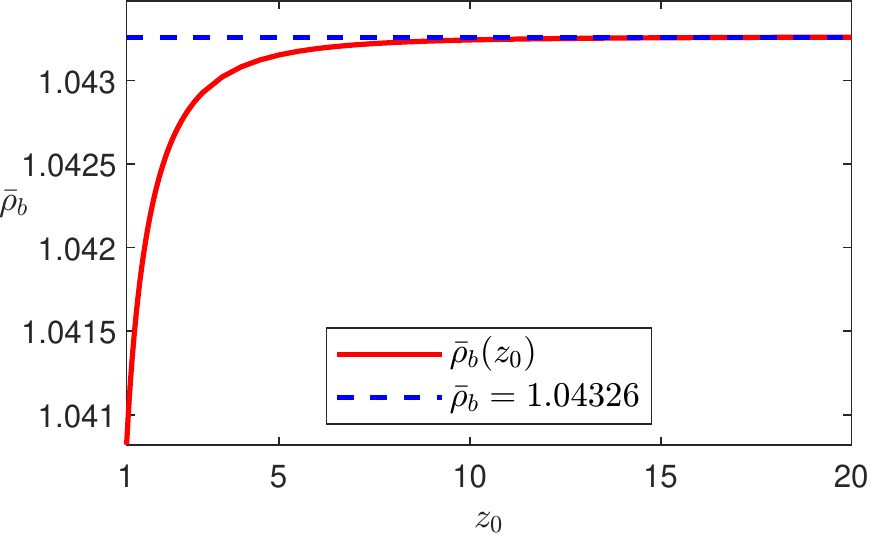}
    }
    \subfigure[]{
    \includegraphics[width=0.46\linewidth]{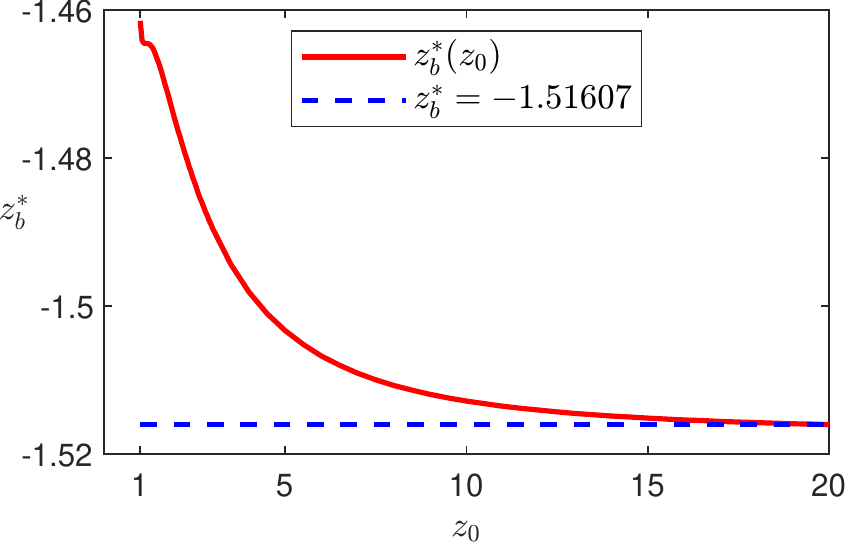}
  }
  \hfill
  \caption[]
  { \textbf{(a)} The red solid curve shows the variations of critical density $\bar{\rho}_{b}$ for a unit sphere as a function of initial position $z_0$ with $\rho_{2}=1.04$. The blue dashed line is the asymptote of the $\bar{\rho}_{b} (z_{0})$ as $z_{0} \rightarrow \infty$. \textbf{(b)} The red solid curve shows the critical point $z^{*}_{b}$ as a function of initial position $z_{0}$. The blue dashed line shows an asymptote of $z^{*}_{b} (z_{0})$. }
  \label{fig:CriticalDensityDifferentPosition}
\end{figure}

We start with the case $L_{\rho}=0$. We explore the dependence of the critical density $\bar{\rho}_b$ for several parameters such as the bottom fluid density $\rho_2$, and the initial position $z_0$. First, as the initial position is closer to the interface, the sphere entrains the light fluid and therefore is harder to levitation.  Figure (a) clearly shows the critical density asymptotically converges as the initial position moving further away from the interface. Similarly, figure \ref{fig:CriticalDensityDifferentPosition} (b) shows the critical point $z^{*}_{b} (z_{0})$ also converges as $z_0 \rightarrow \infty$.

\begin{figure}
  \centering
    \subfigure[]{
    \includegraphics[width=0.46\linewidth]{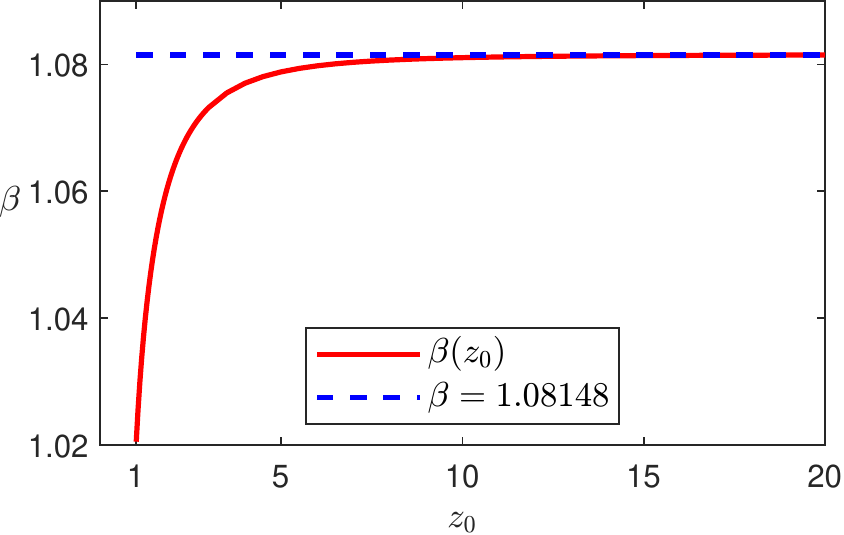}
    }
    \subfigure[]{
    \includegraphics[width=0.46\linewidth]{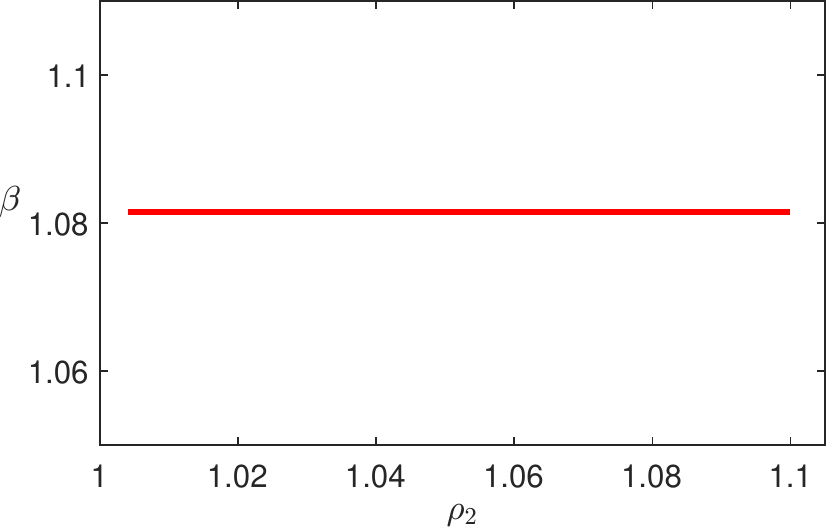}
  }
  \hfill
  \caption[]
  { \textbf{(a)} The red solid line shows the variation of the non-dimensionless parameter $\beta$ as a function of the initial position $z_{0}$.  The blue dashed line is the asymptote of the $\beta (z_{0})$ as $z_{0} \rightarrow \infty$. \textbf{(b)} The figure shows that non-dimensionless parameter $\beta$ is independent of the bottom fluid density $\rho_{2}$.  }
  \label{fig:BetaDifferentPositionRho}
\end{figure}

To further investigate the criterion \eqref{eq:DegenerateCritialPoint}, we take the  derivative of the potential energy equation \eqref{eq:DeltaPe} with respect to $z_b$ and setting it to zero, which yields
\begin{equation}\label{eq:DzDeltape}
\pi(\rho_2-1) \partial_{z_{b}}\left( \int_{r^{*} (z_{b})}^{\infty} r z(r)^2  \mathrm{d} r- \int_{z^{*} (z_{b})}^{0} r(z)^2 z  \mathrm{d} z \right) + \frac{4\pi}{3}(\rho_{b}  -1) =0.
\end{equation}
The second equality in equation \eqref{eq:DeltaPe} becomes
\begin{equation}
\label{eq:Dz2Deltape}
\pi(\rho_2-1) \partial_{z_{b}}^{2}\left( \int_{r^{*} (z_{b})}^{\infty} r z(r)^2  \mathrm{d} r- \int_{z^{*} (z_{b})}^{0} r(z)^2 z  \mathrm{d} z \right)=0.
\end{equation}
Numerically solving the above equation yields the critical position $z_{b}=z_{b}^{*}$. Substituting it back to equation \eqref{eq:DzDeltape} gives the critical density
\begin{equation}
  \label{eq:rho1Andrhob}
\bar{\rho}_{b}= (\rho_{2}-1)\beta+1,\quad \beta = \frac{3}{4} \left. \partial_{z_{b}}\left( \int_{r^{*} (z_{b})}^{\infty} r z(r)^2  \mathrm{d} r- \int_{z^{*} (z_{b})}^{0} r(z)^2 z  \mathrm{d} z \right)  \right|_{z_{b}=z_{b}^{*} },  
\end{equation}
where $\beta$ can be numerically computed. We call attention to three important properties of $\beta$. First,  $\beta$ is dimensionless. Second, as demonstrated in figure \ref{fig:BetaDifferentPositionRho} (a), $\beta$ increases as the sphere initial position $z_{0}$ increases. Third, $\beta$ is independent of $\rho_{2}$ and $\rho_{b}$, which is verified in the figure \ref{fig:BetaDifferentPositionRho} (b).  Also note that in this model the depth $z_b^*$ is independent of the sphere density as is evident from equation \eqref{eq:Dz2Deltape}.

Equation \eqref{eq:rho1Andrhob} shows that $\bar{\rho}_{b}$ linearly depends on $\rho_2$ , which closely resembles equation \eqref{eq:LRExperiment} from the experiments with slightly different coefficients. Additionally, equation \eqref{eq:rho1Andrhob} indicates that the difference between the critical sphere density and the bottom fluid density increases proportionally with the density differences in the fluid and also with the value of $\beta-1$ where $\beta > 1$. Therefore we have equality, namely $\bar{\rho}_s=\rho_2 $ iff $\rho_2=\rho_1$.

\begin{figure}
  \centering
    \includegraphics[width=0.46\linewidth]{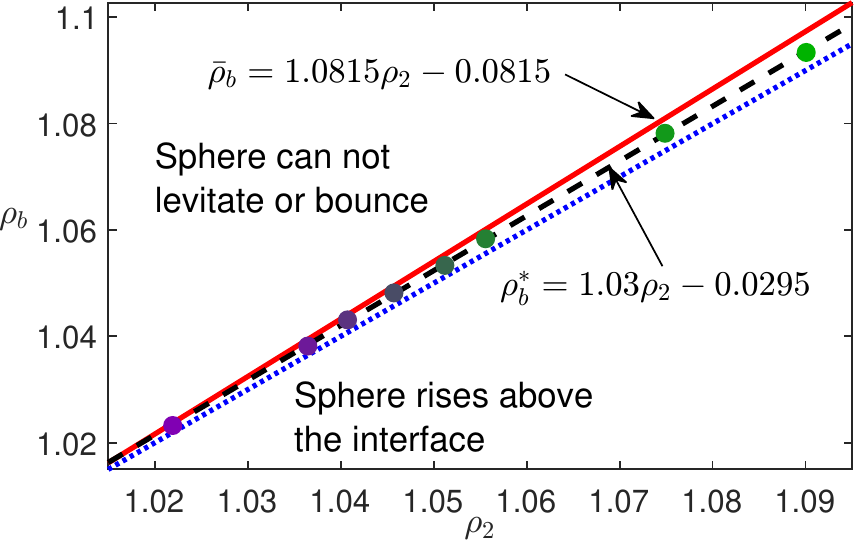}
  \hfill
  \caption[]
  {  Comparison of experimental results and theoretical prediction. The blue dotted line represents a line of slope 1 and the red solid lines are the critical sphere densities obtained by solving equation \eqref{eq:DegenerateCritialPoint} numerically with $z_{0}=20$. The coordinates of the color dots are the non-dimensionalized critical density triplet $(1, \rho_{2}, \rho_{b})$ from table \ref{tab:ExperimentalResults}. The black dashed line is the linear regression \eqref{eq:LRExperiment} of the experimental data. All dots are bounded by the red and blue lines.  }
  \label{fig:CompareExperiment}
\end{figure}

Now, we are ready to compare our numerically obtained values with those obtained from our experiments, which are presented in figure \ref{fig:CompareExperiment}. When $z_{0}=20$, $\bar{\rho}_{b}$  can be expressed as
\begin{equation}\label{eq:rho1AndrhobZ20}
\bar{\rho}_{b}=1.0815 \rho_{2}-0.0815.
\end{equation}
We have two comments about this formula. First, $\bar{\rho}_b$ provided in the above equation shows a relative difference $\frac{\bar{\rho}_{b} (\rho_{2}) -1}{\rho_{b}^{*} (\rho_{2}) -1}-1 \leq 0.043$ for all $\rho_2 \in [1,1.1]$, which is a great improvement compared with the criterion proposed in article \cite{camassa2008brachistochrones}.  Second, the experimental values of critical density from table \ref{tab:ExperimentalResults} and the linear regression \ref{eq:LRExperiment} consistently fall inside  theoretical critical window $[\rho_2,\bar{\rho}_b]$ within the experimental parameter regime $1<\rho_{2}\leq 1.1$.  Figure \ref{fig:CompareExperiment} demonstrates the sphere density $\bar{\rho}_{b}$ obtained from the energy curve constitutes an upper bound for the density $\rho^{*}_{b}$ in the critical density triplet. This behavior is very reminiscent of a Van-der-Waal's type pressure-volume curve \cite{kondepudi2008introduction}.

\subsection{Non-zero density transition layer thickness}
Now, let us switch the attention to the case with non-zero density transition layer thickness. With a similar procedure, we have
\begin{equation}  \label{eq:rho1AndrhobFiniteThickness}
  \begin{aligned}
      &\bar{\rho}_{b}= (\rho_{2}-1)\beta+1_{},\\
  &\beta = \frac{3}{4} \left. \lim\limits_{N\rightarrow \infty}  \sum\limits_{n=1}^N  \frac{  \rho_{I,n}-\rho_{I,n-1} }{\rho_{2}-1}  \partial_{z_{b}}\left( \int_{r_{n}^{*} (z_{b})}^{\infty} r z_{n}(r)^2  \mathrm{d} r- \int_{z_{n}^{*} (z_{b})}^{0} r_{n}(z)^2 z  \mathrm{d} z \right)  \right|_{z_{b}=z_{b}^{*} },  
  \end{aligned}
\end{equation}
where $z_{b}^{*}$ is the location for the summation reaches the minimum value,  which also solves the equation
\begin{equation}  
 \left. \lim\limits_{N\rightarrow \infty}  \sum\limits_{n=1}^N  \frac{  \rho_{I,n}-\rho_{I,n-1} }{\rho_{2}-1}  \partial_{z_{b}}^{2}\left( \int_{r_{n}^{*} (z_{b})}^{\infty} r z_{n}(r)^2  \mathrm{d} r- \int_{z_{n}^{*} (z_{b})}^{0} r_{n}(z)^2 z  \mathrm{d} z \right)  \right|_{z_{b}=z_{b}^{*} }=0.  
\end{equation}
According to equation \eqref{eq:InitialDensityProfile Nondimensional}, we have $\rho (z_{i})-\rho (z_{j})= (\rho_{2}-1)f (z_{i},z_{j})$ for $z_{i},z_{j} \in [-\frac{L_{\rho}}{2}, \frac{L_{\rho}}{2}]$ where $f (z_{i},z_j)$ is independent of $\rho_2$. Hence, the dimensionless parameter $\beta$ defined in equation \eqref{eq:rho1AndrhobFiniteThickness} is independent of $\rho_2$, and only depends on $L_p$ and the initial position of the sphere.

The layer thickness $L$ in section \ref{sec:Experiment} is measured as the distance between the points of 10\% and 90\% change in density. For the density profile provided in equation \eqref{eq:InitialDensityProfile Nondimensional}, we have $L\approx 0.6084L_p$. The layer thickness presented in table \ref{tab:ExperimentalResults} is around $L= \frac{0.85 cm}{0.25 cm} \approx 3.4$. Hence, we numerically evaluate $\beta$ in equation \eqref{eq:rho1AndrhobFiniteThickness} with  $L_{p}=3.4/0.6084 \approx 5.6$ and obtain
\begin{equation}
  \bar{\rho}_{b}=1.0289\rho_{2}-0.0289.
\end{equation}
This estimation shows a relative difference $\frac{\bar{\rho}_{b} (\rho_{2}) -1}{\rho_{b}^{*} (\rho_{2}) -1}-1 \leq 0.0078$ for all $\rho_2 \in [1,1.1]$.

\begin{figure}
  \centering
    \includegraphics[width=0.46\linewidth]{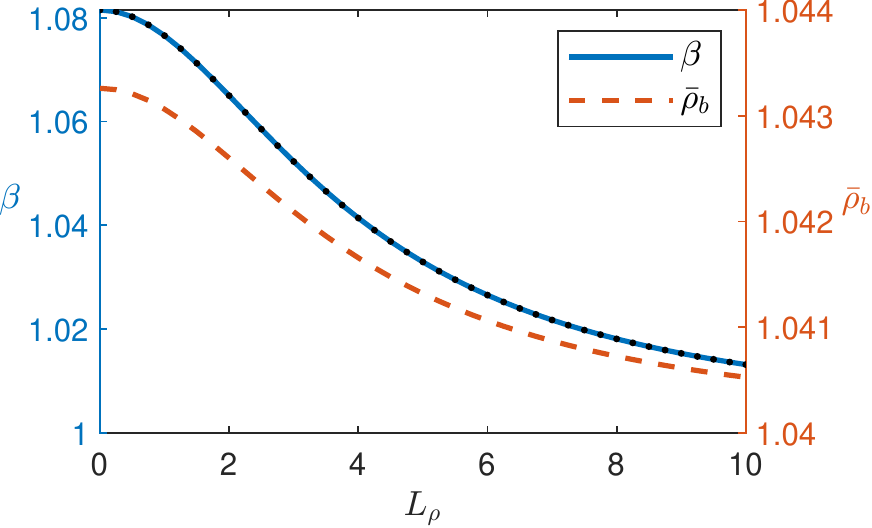}
  \hfill
  \caption[]
  {The blue solid line and red dashed line show  the variation of the non-dimensionless parameter $\beta$ {\color{blue}(left axis)} and estimated critical sphere density $\bar{\rho}_{b}$ {\color{red}(right axis)} provided in equation \eqref{eq:rho1AndrhobFiniteThickness} as functions of $L_{\rho}$, respectively. The parameters are $\rho_2=1.04$, $z_{0}=20+L_{\rho}/2$. For convergence, we need the number of artificial layers $N$ to increase as $L_{\rho}$ increases. We use $N=30$ when $L_{\rho}=1$, and $N=80$ when $L_{\rho}=10$. To show $\beta$ is independent of $\rho_2$, we repeat the simulation for $\rho_2=1.05$ and then plot the resulted $\beta$ with black dots which is fully overlapped with the solid blue curve for the case $\rho_{2}=1.04$. }
  \label{fig:CriticalDensityDifferentLayerThickness}
\end{figure}

Figure \ref{fig:CriticalDensityDifferentLayerThickness} shows the $\beta$ and  $\bar{\rho}_{b}$ decreases as the layer thickness increases which qualitatively captures the trend observed in experiments. We have two remarks. First, one can observe this trend with 3 fluid layers in the simulation. Second, the decreasing rate of $\bar{\rho}_{b}$ with respect to the layer thickness is relatively larger than the experimental observation: In the experiment, the non-dimensionalized $\rho^{*}_{b}$ changes around $0.0002$ as the non-dimensionalized layer thickness $L$ increases from $4$ to $6$, while, figure \ref{fig:CriticalDensityDifferentLayerThickness} shows $\bar{\rho}_{b}$ changes $0.0004$ as $L_{\rho}$ increase from $4/0.6084$ to  $6/0.6084$. The difference in the change rate is because our potential energy-based criterion doesn't consider the complex nature of the fluid flow, for example, the viscous fluid layer around the body. However, even for the large $L_{\rho}$, $\bar{\rho}_{b}$  is an accurate estimation with less than 1\% relative difference which is better than the estimation provided in equation \eqref{eq:rho1AndrhobZ20} under the zero layer thickness assumption.

\section{Conclusion and discussion}
\label{sec:discuss}

We have studied the constraints of the fluid and sphere densities for producing a bouncing or levitation when a rigid sphere falls in a two-layer stratified fluid. Experiments focus on cases with relatively high Reynolds numbers (between 20 and 450) and sharply stratified fluid ($h/a<4$). We explore the critical density triplet $(\rho_1,\rho_2,\rho^{*}_b)$ as defined in section \ref{sec:intro} with experimental and theoretical method. The main results are summarized as follows.

First, experiments shows that the increasing of fluid density transition layer decreases the difference between $\rho_{2}$ and $\rho_{b}^{*}$. Second, when the relative fluid density transition layer thickness $h/a$ is around $3 \sim 4$, the linear regression of the experimental shows that the dimensionless ratio $\rho_{b}^{*}/\rho_{1}$ increases linearly as $\rho_{2}/\rho_1$ increases. Third, based on the monotonicity of the potential energy curve, we identified a critical sphere density $\bar{\rho}_{b}$ which could be the estimation of $\rho^{*}_{b}$.
With the zero layer thickness assumption, the estimation $\bar{\rho}_b$ is an upper bound of the experimental measured $\rho^{*}_{b}$ with less than 0.043 relative difference $\frac{\bar{\rho}_{b} (\rho_{2}) -\rho_{1}}{\rho_{b}^{*} (\rho_{2}) -\rho_{1}}-1$ within the experimental parameter regime $\rho_{2}/\rho_{1} \in [1, 1.1]$. Next, we demonstrated that $\bar{\rho}_{b}$ decreases as the layer thickness increases. When the layer thickness matches the experimental value $\frac{h}{a}\approx 3.4$, we obtain an more accurate estimation $\bar{\rho}_b$ which has a 0.0078 relative difference within the same experimental parameter regime, which is a great improvement compared with the estimation proposed in article \cite{camassa2008brachistochrones}.

Future research include a number of directions. 
First,  the fluid layer surrounding the sphere could play a role in the settling dynamics \cite{abaid2004internal,camassa2012stratified}. In particular, the work using a vertically towed fishing line in stratification \cite{camassa2012stratified} provides a starting point towards estimating the size of this boundary layer.  We expect that including the viscous drag from the fluid layer into the model could yield a more accurate prediction of the critical densities. 
Second, many articles numerically studied the particles settling in an unbounded  stratified fluid   \cite{doostmohammadi2014numerical,torres2000flow,deepwell2021settling}, while fewer studies have addressed the case with sharply stratified fluid and relative high Reynolds numbers. 
We plan to investigate the complicated dynamics of the momentary levitation discussed in this paper using a direct numerical simulation of the Navier-Stokes equation and understand the effect from the rigid boundaries.
Third, we are interested in generalizing the theory to a dual problem, namely the rise of droplets in a sharply stratified fluid \cite{mandel2020retention,shaik2020drag,farhadi2022passage}, which is important in the studying of the oil spill \cite{adalsteinsson2011subsurface,camassa2016optimal}.

\section{Acknowledgements}
 We acknowledge funding received from the following National Science Foundation Grant Nos.:DMS-1910824; and Office of Naval Research  Grant No: ONR N00014-18-1-2490. Partial support for Lingyun Ding is gratefully acknowledged from the National Science Foundation, award NSF-DMS-1929298 from the Statistical and Applied Mathematical Sciences Institute.

 \section{Appendix}
\label{sec:appendix}

\subsection{Numerical Method}
In this section, we document the details of the numerical calculation of the drift and reflux contributions to the potential energy and associated issues. 

As the sphere penetrates the interface and deforms it, there is considerable stretching of the mesh in the region around the sphere, due to the potential nature of the flow.  The uniform mesh on the interface can not resolve the dynamics efficiently. Hence, for simplicity, we adopt a non-uniform mesh, which takes the parameterization,
\begin{equation}
 x(s)=
\begin{cases}
  0&s=0\\
  e^{\frac{s}{r_{1}}} & 0<s \leq r_{1} \\
  k_{1}(s-r_{1})+1 & r_{1}<s
\end{cases}, \quad 
y(s)=0, 
\end{equation}
where $r_1$, $k_1$ are constants selected to resolve the interface evolution profile, which varies for different initial position of the sphere and the duration of the evolution. The mesh points cluster exponentially at the neighborhood of zero, and distributed uniformly when they are far away from zero. 
The exponential profile of the initial mesh in the immediate vicinity of the particle provides a high density of meshes where the stretching is maximum.

A fourth-order explicit Runge-Kutta method with typical step size $\Delta t=10^{-3}$ was used to compute the time evolution of the interface region as the sphere moved through the layers by solving the initial value problem with the velocity field provided in equation \eqref{eq:potentialFlow3D}.

We approximate the interface by the cubic spline with the boundary condition ``not-a-knot'' to ensure 4-th order accuracy in the interface tracking stage. The point of zero Lagrangian displacement $(r^{*} (z_{b}),0)$ is calculated by solving the root of the spline function.

The integrals in equation \eqref{eq:DeltaPe} are evaluated by the trapezoidal rule. To achieve higher accuracy, one can adopt the spline-based quadrature rules described in \cite{zhang2019lagrangian,sommariva2009gauss}. The potential energy as a function of the sphere position is approximated by a 5-th order spline function. Since differentiation could introduce unexpected oscillations when the data is not smooth enough, instead of solving $\partial^{2}_{z_{b}}P (z_{b}^{*})=0$ for the critical point $z_{b}^{*}$, we calculate $z_{b}^{*}$ by finding the minimum value of $\partial_{z_{b}}P$.


We verified that all numerical results were not sensitive to an increase of either spatial or temporal resolution, therefore establishing the convergence of the numerical scheme.

\bibliographystyle{elsarticle-harv}


\end{document}